# PNAS

**Main Manuscript for**

Surface Block Identity Controls Transport of Symmetric Diblock Copolymer Through Nanopores

Sang Yup Lee[1,3,4], Tae-Young Heo[1,4], Uiseok Hwang[1], Théophile Ienn[2], Julien Bernard[2], Robert A Riggleman[1,*], and Daeyeon Lee[1,*]

[1]Department of Chemical Biomolecular Engineering, University of Pennsylvania, Philadelphia, PA, 19104, United States.

[2]Université de Lyon, CNRS, Université Claude Bernard Lyon 1, INSA Lyon, Université Jean Monnet, UMR 5223, Ingénierie des Matériaux Polymères F-69621 Cedex, France

[3]KU-KIST Graduate School of Converging Science and Technology, Korea University; Seoul 02841, Republic of Korea

[4]These authors contributed equally to this work.

*To whom correspondence may be addressed. Email: rrig@seas.upenn.edu and daeyeon@seas.upenn.edu.

**This PDF file includes:**

    Main Text
    Figures 1 to 4
    Tables 1 to 1

**Abstract**

Understanding how polymer architecture governs transport through nanopores is essential for nanocomposite fabrication, membrane design, and polymer upcycling. However, the effect of the nanoscale structure of copolymers on chain transport through nanoporous media remains poorly understood. In this study, we demonstrate that simply inverting the surface orientation of lamellar poly(styrene-block-2-vinylpyridine) (PS-b-P2VP) diblock copolymers, composed of two monomers with strongly contrasting affinities for $SiO_2$, at the entrance of nanoporous silica significantly alters the kinetics of capillary rise infiltration. Using in situ spectroscopic ellipsometry, we find that infiltration of symmetric PS-b-P2VP into silica nanoparticle ($SiO_2$ NP) packings is significantly faster when the P2VP domain is the top layer of the film and first contacts the nanoparticles, compared to when the PS domain is the top layer. Coarse-grained molecular dynamics simulations reveal that this difference originates from block-specific adsorption pathways that reorganize the nanophase structure around nanoparticles: P2VP-first infiltration forms thin adsorbed layers that drive PS into the pore interiors, generating continuous interfacial pathways that enable rapid, interface-mediated transport. In contrast, PS-first infiltration produces thicker P2VP layers that isolate PS domains and disrupt pathway connectivity, forcing chains to rely on a slower, connectivity-limited transport mechanism through P2VP-rich interstitial regions. Above the order–disorder transition, or upon silanizing nanoparticles to neutralize surface affinity, the rate difference disappears. These findings demonstrate how the interplay between nanoscale domain configuration and polymer–surface affinity governs infiltration dynamics, providing mechanistic insight into tuning transport in nanostructured block copolymers and guiding the design of high-loading composites, selective membranes, and scalable upcycling pathways for complex waste streams.

**Significance Statement**

Many technologies, from polymer–nanoparticle nanocomposites to porous catalysts for plastic upcycling, rely on long polymer chains penetrating nanometer-sized pores. Yet for copolymers, it has been unclear how nanoscale architecture, beyond their chemistry, controls this transport. Using a lamellar diblock copolymer whose two blocks have strongly different affinities for silica, we show that simply flipping the surface-facing block dramatically changes the infiltration speed. We find that the first block to reach the surface determines how adsorbed layers form and whether continuous interfacial pathways develop for fast motion or fragmented domains force slow, connectivity-limited transport. These results provide a practical design rule to accelerate pore filling in high-loading nanocomposites and mitigate transport bottlenecks in upcycling reactors.

**Main Text**

**Introduction**
The translational dynamics of polymers under nanoconfinement govern processes central to the fabrication of nanocomposites, the fractionation and analysis of (bio)macromolecules, and the catalytic upcycling of plastic wastes (1–7). When polymer chains move through highly confining pores, their dynamics deviate sharply from bulk behavior (8–12). Strong polymer–surface interactions, for example, can immobilize chains at interfaces, thereby reducing the effective pore size and slowing transport (13–17). Conversely, confining entangled polymers can facilitate motion by reducing entanglement density and leading to faster diffusion (8, 9, 11, 18). These insights guide the design of processes to facilitate polymer infiltration for nanocomposite preparation and to reduce transport limitations in polymer upcycling reactions using porous catalysts, where large polymer chains must penetrate catalyst pores to enable efficient depolymerization.

Copolymers introduce a new level of complexity. They account for a large proportion of commercially used polymers, in which tailored monomer combinations result in tunable properties (19, 20). Also, biopolymers consist of diverse monomeric building blocks, making sequence heterogeneity a likely key factor in their transport through nanopores (5, 21). Moreover, copolymers such as ethylene-vinyl alcohol (EVOH) present in multilayer packaging waste streams pose significant challenges for separation and upcycling (22–24). Thus, understanding how copolymer composition and sequence affect their behavior under confinement is crucial. Recent experiments have indeed shown that statistical copolymers comprised of strongly and weakly interacting monomers exhibit nonmonotonic infiltration dynamics as a function of composition (25). This trend cannot be explained solely by bulk viscosity or glass-transition shifts and is likely strongly influenced by the conformation of chains on the pore surface and by how surface-adsorbed chains interact with other moving chains. Such results suggest that translational dynamics under confinement are sensitive to monomer sequence and segment identity (25–27).

Block copolymers provide a unique platform for probing how chain architecture and segment sequence influence confined polymer dynamics (19, 20). Their ability to self-assemble into ordered nanostructures enables precise control over interfacial composition and has been exploited to direct nanoparticle organization, template membranes, and fabricate porous materials (28–30). However, it remains unclear how such a prearranged domain structure affects nanopore infiltration. Here, we investigate a model lamellar poly(styrene-*block*-2-vinylpyridine) (PS-*b*-P2VP) diblock copolymer, composed of two monomers with weak (PS) and strong (P2VP) affinities for $SiO_2$, to probe how the domain arrangement at the pore entrance influences their transport through the pores. We find that simply inverting the block at the free surface, without altering its chemistry, composition, or morphology, results in strikingly different infiltration rates. This sensitivity of macroscopic transport to nanoscale domain arrangement reveals a previously unrecognized mechanism for controlling polymer infiltration under confinement. To uncover the molecular origins of this phenomenon, we rely on molecular dynamics (MD) simulations and *in situ* infiltration experiments. Simulations reveal how the block that first encounters the nanoporous medium during infiltration organizes the adsorbed layers, thereby creating either continuous or obstructed pathways, which significantly affect the dynamics of subsequent infiltrating chains. Inducing infiltration above the order–disorder transition temperature, or passivating nanoparticles to equalize polymer–nanoparticle interactions, renders the two infiltration cases effectively equivalent. Our findings suggest new strategies for tuning translational dynamics of block copolymers in nanocomposite fabrication, reducing transport limitations in catalytic upcycling of plastics, and optimizing selective transport in porous materials.

## Results

**Lamellar Surface Termination Controls Nanopore Infiltration.**
Two types of lamellar PS-*b*-P2VP films with different top layers are prepared to investigate surface termination-dependent infiltration dynamics of symmetric diblock copolymer. PS-*b*-P2VP is an ideal model block copolymer because it comprises two monomers with markedly different affinities for $SiO_2$. P2VP interacts strongly with $SiO_2$ via hydrogen bonding, whereas the interaction between $SiO_2$ and PS is relatively weak. The diblock copolymer ($M_n$ = 16.2 kg mol$^{-1}$; $f_{PS}$=0.52) used in this study has a glass transition temperature ($T_g$) of ~96 °C and an order–disorder transition temperature ($T_{ODT}$) of ~167 °C (*SI Appendix*, Fig. S4). Annealing a ~650 nm-thick disordered film at 150 °C, between $T_g$ and $T_{ODT}$, induces microphase separation into a lamellar structure with the domain spacing of 14 nm. Because of their higher polarity, P2VP blocks preferentially wet the silicon (Si) substrate, whereas PS blocks segregate to the air interface, creating what we call PS-on-top films.

To invert the surface termination of the lamellar film, the PS-on-top film is coated with a hydrophilic poly(vinyl alcohol) (PVA) layer and annealed at 140 °C. A similar approach has previously been used to induce a more polar block to the top surface in block copolymer films with microphase-separated domains (31–33). Under such a condition, the lamellae reorganize to place P2VP at the top of the film (P2VP-on-top). After dissolving the PVA in water, P2VP-on-top films are obtained (Fig. 1A). Contact angle measurements and spectroscopic ellipsometry of Au-labelled lamellar films confirm the two types of orientations, which we summarize in *SI Appendix,* Fig. S3 and Fig. S5.

Random packings of $SiO_2$ NPs (22 nm diameter) are placed on top of both PS-top and P2VP-top films to study the dynamics of capillary rise infiltration (CaRI). The packing has an average thickness of 622 nm and a pore size of ~ 6.6 nm ($D_{pore}$ ≈ 0.3 × $D_{NP}$), respectively. The average pore size of the NP packing has been previously measured experimentally using ellipsometric porosimetry (34). Upon annealing above $T_g$, the block copolymer infiltrates the NP layer, filling the interstitial voids (Fig. 1B). We use *in situ* ellipsometry to track the time-dependent thickness of the polymer-filled NP layer (Fig. 1C). Infiltration kinetics are quantified using the infiltration time ($t_{inf}$), defined as the time required to fully fill a 622 nm-thick NP layer.

The infiltration time strongly depends on the configuration of the lamellar film; that is, the domain initially in contact with the $SiO_2$ NP packing significantly influences the infiltration dynamics. At 150 °C, the P2VP-on-top film reaches saturation in 429 min, whereas the PS-on-top film requires 686 min (Fig. 1D). Even at an elevated temperature of 160 °C, the P2VP-on-top film fills the NP packing significantly faster than the PS-on-top film. The difference in infiltration dynamics develops from the early stage of infiltration, as shown in Fig. 1E. Given that the two cases are based on the same symmetric diblock copolymer, it is unlikely that its bulk viscosity or glass transition temperature is responsible for the difference in the infiltration dynamics. As a point of reference, infiltration dynamics of PS and P2VP homopolymers, as well as statistical copolymers of the same composition (2VP:S = 50:50), show significantly faster infiltration kinetics compared to the two cases with the symmetric diblock copolymers, indicating that chemical composition alone does not account for the observed disparity (*SI Appendix*, Fig. S6).

**Molecular Simulations Reveal Domain Connectivity as the Origin of Transport Differences.**
To understand the observed difference in infiltration rates between the two lamellar configurations (Fig. 2A and B), we perform MD simulations using the Martini coarse-grained (CG) model (35, 36) (See *SI Appendix* for the detailed description of MD simulations). MD simulations of CaRI are carried out at 130 °C for 1 μs. Consistent with the experimental observation, the P2VP-on-top system exhibits significantly faster infiltration kinetics than the PS-on-top system as shown in Fig. 2C and D. To analyze infiltration kinetics, we track the infiltration depth, defined as the z-position at which the polymer density exceeds 100 kg/m$^3$, as a function of time. As shown in Fig. 2D, the infiltration depth increases more rapidly in the P2VP-on-top system from the early stage at 130 °C.

Calculated over the 200–800 ns range (excluding the initial wetting stage), the average infiltration rates are 0.019 nm/ns for P2VP-on-top and 0.011 nm/ns for PS-on-top, indicating that the P2VP-on-top system infiltrates almost 1.7 times faster than the PS-on-top system under these conditions.

To elucidate the origin of the difference in the infiltration dynamics, we analyze the arrangement of diblock copolymers in the interstices of the NP packing during infiltration. Cross-sectional snapshots reveal markedly different morphology evolution for the two configurations during infiltration (Fig. 3). In the P2VP-on-top system (Fig. 3A), P2VP blocks (blue spheres) rapidly adsorb onto and tightly wrap around $SiO_2$ NPs within the first 200 ns. In the PS-on-top case, surface-exposed PS blocks initially contact and adsorb onto the $SiO_2$ NPs (Fig. 3B). Between 200 and 600 ns, however, major reorganization occurs as P2VP blocks gradually migrate to the NP surface and replace the PS blocks on $SiO_2$ NPs due to the stronger affinity of P2VP for the $SiO_2$ surface (37), mediated by hydrogen bonding between the pyridine nitrogen and surface silanol groups (*SI Appendix*, Fig. S12).

Single-chain analysis reveals distinct interfacial rearrangement behaviors. In the P2VP-on-top system (Fig 3C), the number of contacts between 2VP units and $SiO_2$ NP increases sharply at early times and remains constant thereafter, indicating rapid adsorption. In contrast, in the PS-on-top system, PS segments initially occupying the NP interface are gradually replaced by P2VP blocks. We further analyze the radial density distributions of P2VP and PS segments around the bottommost $SiO_2$ NPs, focusing on the final 10 ns range (990–1000 ns). In the P2VP-on-top case (Fig. 3D), a sharp P2VP density peak at ~3 nm from the NP surface indicates the formation of a thin P2VP layer. In contrast, the PS-on-top case exhibits a broader P2VP distribution extending out to ~5 nm with only the ends of P2VP segments attached to the NP. This difference arises from distinct adsorption pathways; in the P2VP-on-top case, P2VP segments adsorb through immediate interactions between P2VP and $SiO_2$ NPs, whereas in the PS-on-top case the NP surface is initially covered by PS segments and later replaced by P2VP segments through molecular rearrangement.

The difference in the molecular organization of diblock copolymers around $SiO_2$ NPs results in significantly different domain connectivity within the porous medium. Figures 4A and 4B show representative domain network structures for the two top-interface configurations at 400 ns. In the P2VP-on-top system (Fig. 4A), PS domains form a continuous, percolated morphology across the interstitial regions between nanoparticles. In contrast, the PS-on-top system (Fig. 4B) exhibits fragmented PS domains separated by interparticle regions highly enriched in P2VP, resulting in isolated PS domains within the nanoparticle packing. These distinct morphologies suggest fundamentally different connectivity of transport pathways for the diblock copolymer chains.

To quantify these connectivity differences, we perform cluster analysis of PS domains (Fig. 4C). PS clusters are defined as assemblies of PS beads within a cutoff distance of 0.6 nm that contain more than 120 PS beads. Owing to the lamellar morphology, both lamellar PS-*b*-P2VP films initially exhibit three PS domains. In the P2VP-on-top configuration, the number of PS clusters rapidly decreases to one during the early stage (0–100 ns), indicating the formation of a single percolated PS domain. In contrast, the PS-on-top configuration maintains three PS clusters up to ~200 ns, demonstrating that domain separation persists during this time window. After 700 ns, the number of clusters in the PS-on-top configuration increases dramatically, indicating the formation of several isolated domains. We believe this is due to the relatively rapid polymer mobility at the infiltration front combined with the slow polymer influx from below, allowing the infiltration front to become more diffuse with more isolated domains.

We obtain further insight by examining the size of the largest cluster, which quantifies the fraction of PS beads in the percolated network (Fig. 4D). In the P2VP-on-top system, this fraction rapidly increases to ~1.0, indicating that nearly all PS beads (~60,000 in total) become interconnected. In contrast, the PS-on-top system remains at a fraction of ~0.6 throughout the simulation window (up to 1000 ns), demonstrating the persistence of multiple, disconnected PS domains. Together, these results demonstrate that the distinct infiltration dynamics originate from differences in domain connectivity established by the surface termination of the lamellar films.

Consistent with prior studies on the bulk (i.e., unconfined) samples of phase-separated diblock copolymers, chain transport is strongly suppressed when domains are isolated but is markedly enhanced when continuous interfaces are present (38). In systems with isolated domains, chains move via activated hopping between domains, whereas continuous interfacial pathways enable rapid diffusion along interfaces (39, 40). Accordingly, the high degree of PS domain connectivity in the P2VP-on-top configuration supports fast, interface-mediated transport, whereas the fragmented domain structure in the PS-on-top case restricts motion and slows infiltration of the diblock copolymer chains upward.

To further assess the impact of domain connectivity on local chain dynamics, we calculate the mobility profiles of polymer blocks along the *z*-direction (Fig. 4E). In the underlying polymer region, the chain mobility is ~1.4 nm/ns. Upon infiltration into the NP-packed region, mobility decreases due to confinement. In the P2VP-on-top case, polymer mobility drops slightly to ~1.0 nm/ns near the gap between NPs. In the PS-on-top case, mobility decreases more substantially, down to ~0.8 nm/ns, and further declines to ~0.5 nm/ns toward the interior of the $SiO_2$ NP-packed region. The significant reduction in the mobility of diblock copolymer chains in the PS-on-top configuration supports the proposed mechanism, in which polymer transport is dominated by connectivity-limited movement and therefore proceeds more slowly.

**High Temperature and Surface Passivation Equalize Transport Pathways**
To understand the importance of the immiscibility of the two blocks and the role of polymer–nanoparticle interactions, we investigate infiltration dynamics at elevated temperature and under conditions of weakened polymer–NP affinity. Raising the temperature to 180 °C, above the order–disorder transition temperature ($T_{ODT}$), eliminates the infiltration-rate difference between P2VP-on-top and PS-on-top films, as observed experimentally (Fig. 5A), demonstrating the important role of microphase separation plays on the slow infiltration dynamics. We also test the importance of the difference in the affinity of the two blocks toward the pore surface by silanized $SiO_2$ NPs at 150 °C (below $T_{ODT}$). Silanization dramatically reduces the number of silanol groups on the $SiO_2$ NP surface, thereby also reducing the hydrogen bonding between 2VP and $SiO_2$. Diblock copolymers infiltrate the $SiO_2$ NP packing much more rapidly, consistent with the results obtained when statistical copolymers of PS-*stat*-P2VP infiltrate silanized $SiO_2$ nanopores (25). More importantly, there is little difference in the infiltration dynamics, highlighting the importance of the different affinities of the two blocks for $SiO_2$ and their subsequent effects on infiltration.

To understand the molecular origin of the convergent behavior induced by high temperature and surface passivation, we perform MD simulations. Consistent with the experimental results, the infiltration rates of the two cases show little difference at 180 °C (Fig. 5B). At 130 °C, the PS-on-top system forms P2VP-rich regions, especially between nanoparticles, which separate PS-rich regions (Fig. 5D, left). Upon increasing the temperature above $T_{ODT}$, P2VP segments retain a stronger affinity than PS but exhibit reduced preferential interaction with the $SiO_2$ surface (SI Appendix, Fig. S12). As a result, both configurations develop similar segment arrangements around the nanoparticles during infiltration (Fig. 5D, center). Enhanced miscibility between PS and P2VP prevents the formation of P2VP-rich domains between nanoparticles. Under these conditions, block–block interfaces are no longer well defined, and interfacial connectivity no longer dominates transport behavior. Consequently, infiltration dynamics become insensitive to the initial surface termination, yielding comparable infiltration rates for P2VP-on-top and PS-on-top systems.

To elucidate the effect of surface passivation, we modify the simulations by changing the NP bead type from one that represents the $SiO_2$ NPs with surface silanol groups to one that models silanized NPs with reduced polymer–NP affinity; see the simulation methods for details. Changing the coarse-grained bead type reduces the difference in interaction strength between NP–P2VP and NP–PS from 1.26 kJ/mol to 0.23 kJ/mol. With this decrease in the polymer–NP affinity, the difference in infiltration behavior between PS-on-top and P2VP-on-top configurations disappears. As the interaction contrast between the nanoparticles and the two blocks decreases, both blocks adsorb to the nanoparticle surface in roughly equal proportions. For surface-passivated $SiO_2$ nanoparticles, the selective replacement of PS segments near the surface by P2VP segments is largely suppressed. As a result, block exchange near the nanoparticle surface is reduced, the formation of P2VP-rich regions between nanoparticles is suppressed, and structural differences between PS-on-top and P2VP-on-top systems disappear, resulting in comparable infiltration pathways and dynamics.

**Discussion**

Our combined experiments and MD simulations show that the infiltration dynamics of symmetric PS-*b*-P2VP diblock copolymers into $SiO_2$ NP packings are governed by the interplay between polymer–NP interaction strength and the surface termination of the lamellar domains. Merely inverting which block is exposed at the free surface significantly alters how polymer chains initially interact with the NPs and evolve under confinement. When the P2VP domain first infiltrates the NP packings, its strong affinity for $SiO_2$ leads to rapid formation of a thin layer that forms percolating domains of PS, resulting in continuous interfaces. Such morphology facilitates the upward flux of polymers through the pores. In contrast, when PS faces the NP packing, the initially PS adsorbs on $SiO_2$ NPs and subsequently P2VP replaces PS on the NP surfaces, generating a thicker interfacial layer that leads to the separation of PS domains. Thus, diblock copolymer chains predominantly use a connectivity-limited transport mechanism to undergo upward motion. At temperatures above the order–disorder transition, where the lamellar morphology is lost and the two blocks become miscible, the infiltration rates for both configurations converge. Likewise, chemical silanization of $SiO_2$, which minimizes the difference in the polymer-NP interactions, eliminates the difference in the infiltration kinetics. Together, these findings demonstrate that the nanoscale arrangement of block copolymers at the pore entrance and the difference in the strength of polymer-NP interactions govern the formation of continuous pathways during infiltration. This insight provides a mechanistic basis for tuning infiltration kinetics through controlled interfacial design, offering practical guidelines for fabricating high-loading polymer-infiltrated nanoparticle films, optimizing confined polymer transport, and rational processing of hybrid materials where chain–surface interactions dictate macroscopic behavior.

**Materials and Methods**

**Materials.** Aqueous dispersion of $SiO_2$ NPs with a diameter of 22 nm (Ludox TM-50) is purchased from Sigma-Aldrich. Poly(vinyl alcohol) (PVA) ($M_w$ = 13,000−23,000 g/mol), toluene, hydrochloric acid (HCl, 37%), and gold(III) chloride trihydrate ($HAuCl_4 \cdot 3H_2O$, >99.9%) are purchased from Sigma-Aldrich. All other chemicals and solvents were used as received without further purification.

**Polymer Synthesis.** Poly(styrene-*b*-2-vinylpyridine) (PS-*b*-P2VP) block copolymer is prepared by reversible addition-fragmentation chain transfer (RAFT) polymerization (see details in *SI Appendix*). The synthesized polymers are characterized by the combination of size exclusion chromatography (SEC, Shimadzu modular system, comprising an SIL-20AC autoinjector, an LC-20AD pump, a DGU-20A degasser, a CTO-20A column oven, an RID-10A differential refractive index detector and a Waters WAT044228 column) and 1H nuclear magnetic resonance spectroscopy (1H NMR). The molecular weight and the overall molecular weight distribution (Đ) are estimated by $H^1$ NMR and from SEC traces using THF as eluant at 40°C and a calibration curve established with polystyrene standards, as shown in *SI Appendix*, Fig. S1. The composition ratio of PS and P2VP blocks in copolymer is determined by relative integration of the peaks from PS and P2VP aromatic rings, as shown in *SI Appendix*, Fig. S2. Polymer characteristics are summarized in Table 1. The block copolymer samples are denoted as Bx, where x is the fraction of 2VP repeat unit (P2VP) in the backbone.

**Preparation of Block Copolymer Films with Lamellar Structure with PS or P2VP as the Top Layer.** Lamellar structures in block copolymer films are prepared through temperature annealing in a vacuum oven. Silicon wafers are cut into approximately 1 cm × 1 cm squares and then rinsed with isopropanol and deionized water. Block copolymer is dissolved in toluene, and the filtered polymer solution is deposited onto a silicon wafer. To prepare lamellar structure with PS as the top layer, the polymer films are annealed at 140 °C for 1 day in a vacuum oven. After annealing, a PS top layer forms due to asymmetric wetting conditions stemming from the lower surface tension of PS relative to P2VP. To prepare lamellar structures with P2VP as the top layer, a PVA layer is deposited on top of the block copolymer film by spin coating, and the block copolymer film is annealed at 140 °C for 1 day in a vacuum oven. The P2VP top layer forms under symmetric wetting conditions, resulting from the lower interfacial tension between PVA and P2VP compared to PVA and PS (See *SI Appendix*). The residual PVA layer is completely removed by sonication for 15 min, as confirmed by FT-IR measurements shown in *SI Appendix*, Fig. S2. The formation of lamellar structures with a PS or P2VP termination layer is verified through a spectroscopic ellipsometer after selectively staining P2VP domains with $AuCl_4$. (See *SI Appendix*)

**Preparation of NP/Polymer Bilayer Films.** To prepare the bilayer films, $SiO_2$ NPs are deposited onto annealed block copolymer films with PS or P2VP as the top layer. $SiO_2$ NP solutions are filtered and then deposited onto the hydrophilic surfaces of polymer films formed by oxygen plasma treatment for 3 sec to prepare a $SiO_2$ NP layer. To eliminate the hydrogen bonding interaction between 2VP and $SiO_2$, the epoxy-modified $SiO_2$ is used, as described in a previous study (25). The epoxy-modified $SiO_2$ NP suspension is deposited using same procedure as mentioned above.

**Characterization of CaRI of Polymers.** Polymer infiltration into the interstices of packed NP films is monitored *in situ* using a J.A. Woollam Alpha-SE spectroscopic ellipsometer. A bilayer sample is adhered to the temperature-controlled stage (Linkam THMS350V) using thermal paste and annealed above the $T_g$ of the polymer. A set temperature (*i.e.,* 150°C and 180°C), heating rates (*i.e.,* 30 °C/min), and hold time for the set point are specified using the Linksys software. Temperature annealing is stopped when there is no change in the dynamic data of psi (ψ) and delta (Δ). The ellipsometry data of ψ and Δ are collected in the wavelength (λ) range of 380−900 nm at an incident angle of 70°. All data processing is conducted using the CompleteEASE software package (41) provided by J.A. Woollam. To capture the time-dependent thickness change during CaRI process, the collected raw data is analyzed by fitting it to the three-layer Cauchy model for $SiO_2$ NP packing layer, polymer-filled NP layer, and polymer layer, respectively. The Cauchy model for each layer is expressed as $n(\lambda) = A + B/\lambda^2 + C/\lambda^4$, where *n, A, B, C,* and λ are the refractive index, the optical constants and the wavelength. To reduce the degree of freedom in model fit, the refractive index of each layer is fixed.

**Molecular Dynamics Simulation.** All-atom MD simulations are performed to parameterize the bonded potentials of the coarse-grained (CG) block copolymer model. The atomistic structures of the PS-*b*-P2VP diblock copolymer and $SiO_2$ NPs are constructed using the CHARMM-GUI Polymer Builder (42) and Nanoparticle Builder (43), respectively. Molecular structures are drawn and exported as MOL2 files using MarvinSketch (ChemAxon) (44), which is used to construct input geometries for subsequent simulations. The CHARMM36 force field (45) is employed to describe all bonded and non-bonded interactions, including polymer–polymer and polymer–NP interactions. To extract bonded parameters for CG mapping, a single PS-*b*-P2VP chain is solvated in hexane, a good solvent for both blocks, to emulate dilute conditions and reduce intrachain entanglement. The system is energy-minimized and equilibrated using NVT and NPT ensembles, followed by a 50 ns production run at 130 °C and 1 atm using a v-rescale thermostat (46) and Parrinello-rahman barostat (47). The simulation time step is set to 2 fs, and all bonds involving hydrogen atoms were constrained using the SHAKE algorithm (48). Post-processing is performed to obtain distributions of bond lengths, angles, and dihedral angles based on the center-of-mass (COM) mapping scheme employed in the CG model (*SI Appendix*, Fig. S13). These distributions are used to derive bonded potentials via Boltzmann inversion and spline fitting, providing input for the CG Martini-based model. All simulations are carried out using GROMACS 2023.3 (49).

**Acknowledgments**

This work was supported by the National Science Foundation (NSF) Process Systems, Reaction Engineering, and Molecular Thermodynamics Program (award no. CBET-1933704), and the National Research Foundation of Korea (NRF-RS-2023-00214066, RS-2024-00463084, and RS-2025-02263336).

26. G. J. Schneider, L. Willner, Chain conformation and structure of adsorbed polymer layers at solid surfaces. *Macromolecules* **43**, 6828–6835 (2010).
27. D. A. Barkley et al., Chain conformation near the buried interface in nanoparticle-filled polymers. *ACS Appl. Mater. Interfaces* **9**, 10875–10886 (2017).
28. T. P. Russell, A. Menelle, J. F. Ankner, G. P. Felcher, Ordering in block copolymer thin films. *MRS Bull.* **27**, 12–18 (2002).
29. S. P. Nunes, K.-V. Peinemann, Membranes based on self-assembled block copolymers: From basic research to large-scale applications. *Angew. Chem. Int. Ed.* **52**, 8150–8167 (2013).
30. J. R. Werber, C. O. Osuji, M. Elimelech, Materials for next-generation desalination and water purification membranes. *Nat. Rev. Mater.* **1**, 16018 (2016).
31. C. M. Bates et al., Polarity-switching top coats enable orientation of sub–10-nm block copolymer domains. *Science* **338**, 775–779 (2012).
32. I. H. Ryu et al., Interfacial energy-controlled top coats for gyroid/cylinder phase transitions of polystyrene-block-polydimethylsiloxane block copolymer thin films. *ACS Appl. Mater. Interfaces* **9**, 17427–17434 (2017).
33. E. Kim et al., A top coat with solvent annealing enables perpendicular orientation of sub-10 nm microdomains in Si-containing block copolymer thin films. *Adv. Funct. Mater.* **24**, 6981–6988 (2014).
34. B. Q. Kim et al., Water-induced separation of polymers from high nanoparticle-content nanocomposite films. *Small* **19**, 2302676 (2023).
35. S. J. Marrink et al., The MARTINI force field: Coarse grained model for biomolecular simulations. *J. Phys. Chem. B* **111**, 7812–7824 (2007).
36. P. C. T. Souza et al., Martini 3: A general purpose force field for coarse-grained molecular dynamics. *Nat. Methods* **18**, 382–388 (2021).
37. S. Gulati, R. P. Quirk, Adsorption of poly(2-vinylpyridine) and polystyrene on silica. *J. Colloid Interface Sci.* **87**, 54–68 (1982).
38. H. Yokoyama, Diffusion of block copolymers. *Mater. Sci. Eng. R Rep.* **53**, 199–248 (2006).
39. S.-H. Choi, T. P. Lodge, F. S. Bates, Mechanism of molecular exchange in diblock copolymer micelles: Hypersensitivity to core chain length. *Phys. Rev. Lett.* **104**, 047802 (2010).
40. J. Lu, F. S. Bates, T. P. Lodge, Remarkable effect of molecular architecture on chain exchange in triblock copolymer micelles. *Macromolecules* **48**, 2667–2676 (2015).
41. J.A. Woollam Co., Inc., *CompleteEASE Software* (J.A. Woollam Co., Inc., Lincoln, NE, USA).
42. J. Lee et al., CHARMM-GUI Polymer Builder for modeling and simulation of synthetic polymers. *J. Chem. Theory Comput.* **17**, 1229–1236 (2021).
43. S. Kim et al., CHARMM-GUI Nanoparticle Builder for modeling and simulation of nanoparticles and their composites. *J. Chem. Theory Comput.* **18**, 479–493 (2022).
44. ChemAxon, *MarvinSketch* 21.17 (ChemAxon Ltd., 2021).
45. J. Huang, A. D. MacKerell Jr., CHARMM36 all-atom additive protein force field: Validation based on comparison to NMR data. *J. Comput. Chem.* **34**, 2135–2145 (2013).
46. G. Bussi, D. Donadio, M. Parrinello, Canonical sampling through velocity rescaling. *J. Chem. Phys.* **126**, 014101 (2007).
47. M. Parrinello, A. Rahman, Polymorphic transitions in single crystals: A new molecular dynamics method. *J. Appl. Phys.* **52**, 7182–7190 (1981).
48. J.-P. Ryckaert, G. Ciccotti, H. J. C. Berendsen, Numerical integration of the Cartesian equations of motion of a system with constraints: Molecular dynamics of n-alkanes. *J. Comput. Phys.* **23**, 327–341 (1977).
49. M. J. Abraham et al., GROMACS: High performance molecular simulations through multi-level parallelism from laptops to supercomputers. *SoftwareX* **1–2**, 19–25 (2015).

**Figures and Tables**

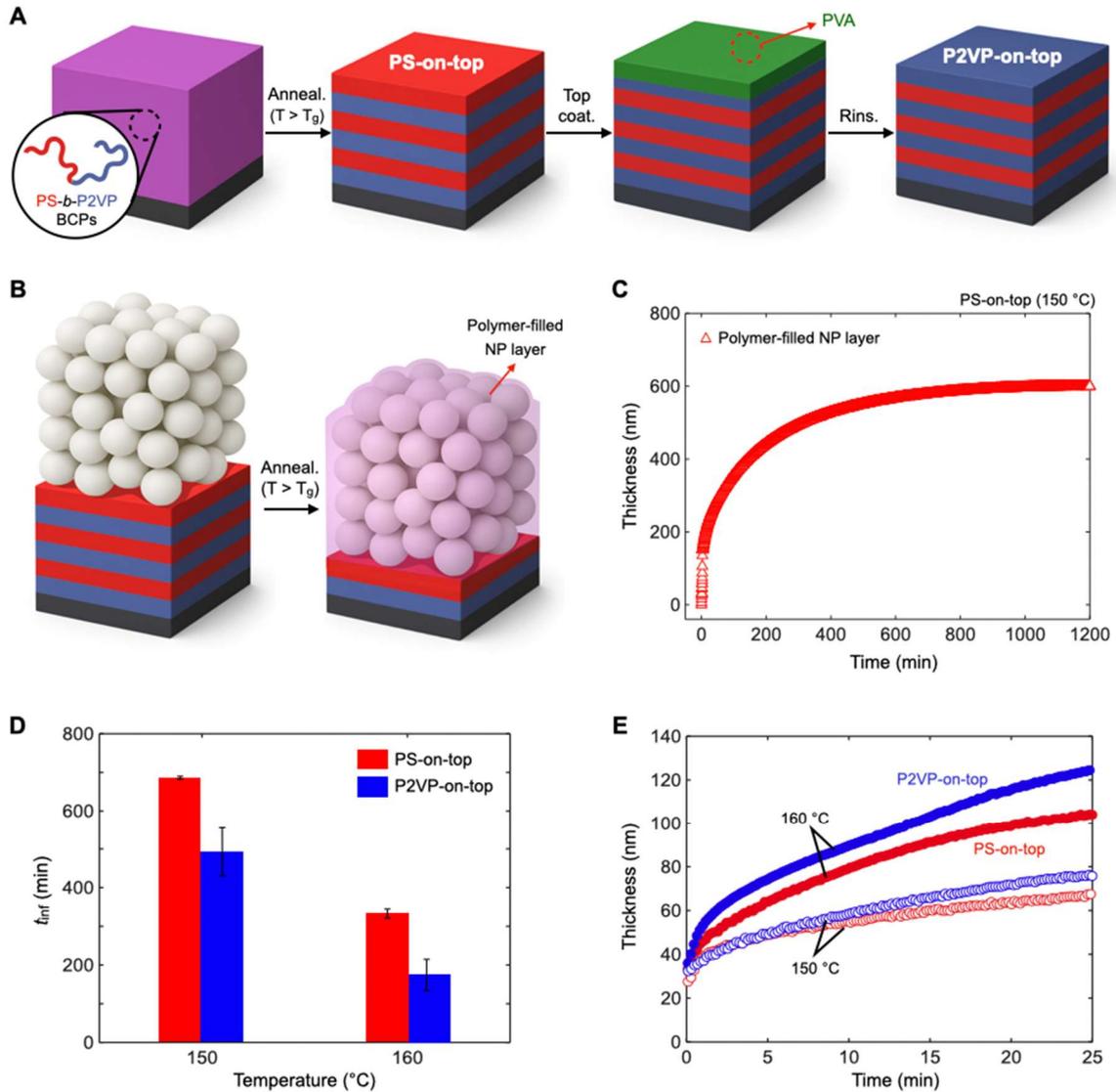

**Figure 1. Difference in the infiltration kinetics depending on the lamellar termination.** (*A*) Schematic illustration of the fabrication process for parallel lamellar films with different top interfaces (PS-on-top and P2VP-on-top). (*B*) Schematic showing the formation of PINFs via CaRI from a PS-*b*-P2VP lamellar film. (*C*) *In situ* ellipsometry measurement showing the thickness evolution of the polymer-filled NP layer (red) during CaRI. (*D*) Histogram of the infiltration time for PS-on-top (red) and P2VP-on-top (blue) films at 150 °C and 160 °C. The infiltration time was extracted from the evolution of the polymer-filled NP layer thickness, following the analysis procedure used in previous CaRI studies (25). (*E*) Early-stage (first 25 min) in-situ thickness evolution of the two lamellar films at 150 °C and 160 °C.

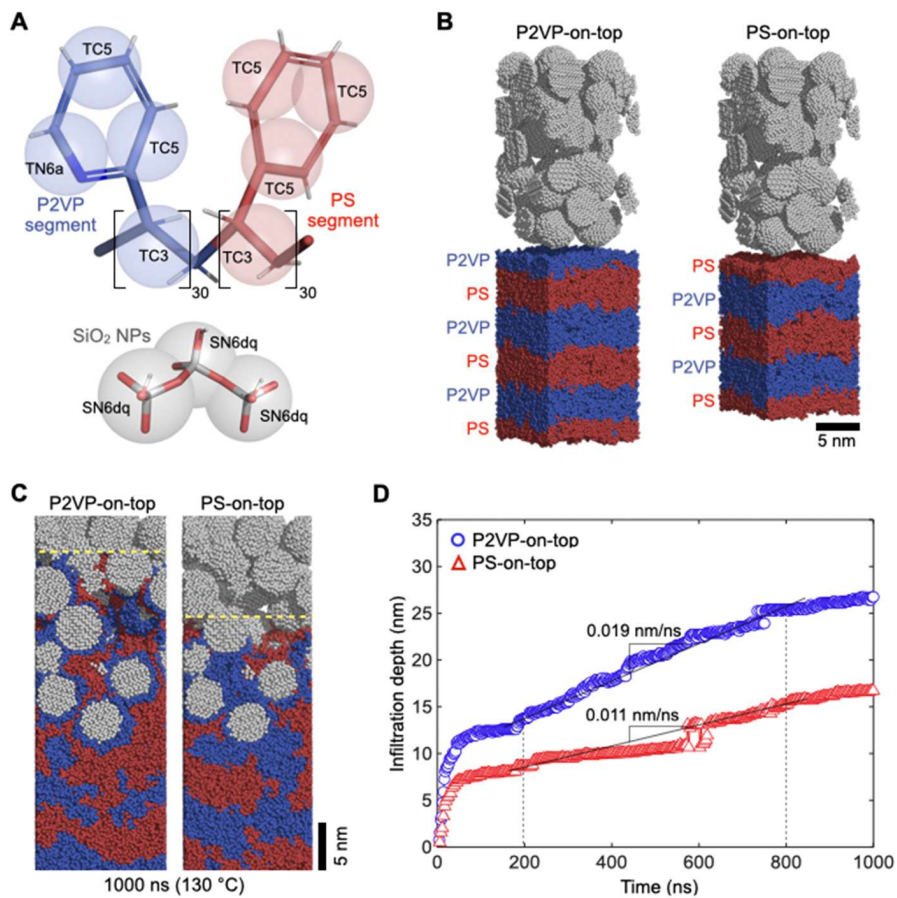

**Figure 2.** Simulation of infiltration behavior for the two lamellar configurations. (A) CG models of the PS-*b*-P2VP diblock copolymer and silica nanoparticle. (*B*) Initial configurations using lamellar films of the block copolymer: P2VP-on-top and PS-on-top. (*C*) Snapshots of the P2VP-on-top and PS-on-top systems at 1000 ns. $SiO_2$ NPs, P2VP segments, and PS segments are represented by gray, blue, and red spheres, respectively. Yellow dashed lines indicate the polymer front. (*D*) Time evolution of infiltration depth, with average infiltration rates calculated between 200 and 800 ns.

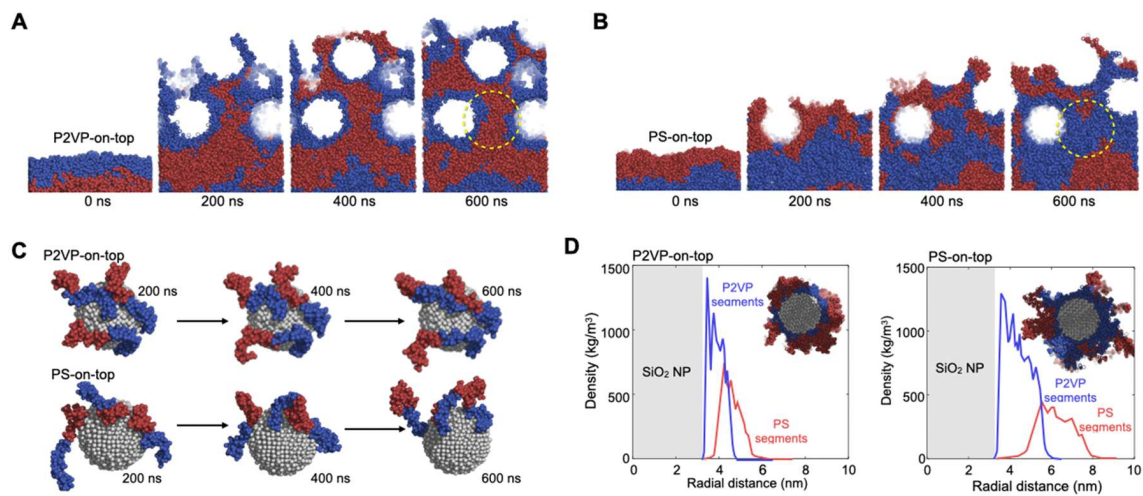

**Figure 3.** Morphology evolution during CaRI of symmetric PS-*b*-P2VP diblock copolymers. (*A* and *B*) Cross-sectional snapshots of P2VP-on-top (*A*) and PS-on-top (*B*) systems at 0, 200, 400, 600, and 800 ns. Only polymer beads within 5 < x < 8 nm are shown; SiO$_2$ NPs are rendered transparent. Yellow dashed circles indicate PS-enriched channels forming during infiltration. (*C*) Snapshots showing the orientation of block copolymer chains around SiO$_2$ NPs for P2VP-on-top (top) and PS-on-top (bottom) systems. Plots show time evolution of the bound fraction between P2VP segments and the nearest SiO$_2$ NP. (*D*) Radial density profiles of P2VP and PS segments around the bottommost SiO$_2$ NP for P2VP-on-top and PS-on-top systems. Shaded region denotes the SiO$_2$ core; P2VP and PS shown in blue and red, respectively.

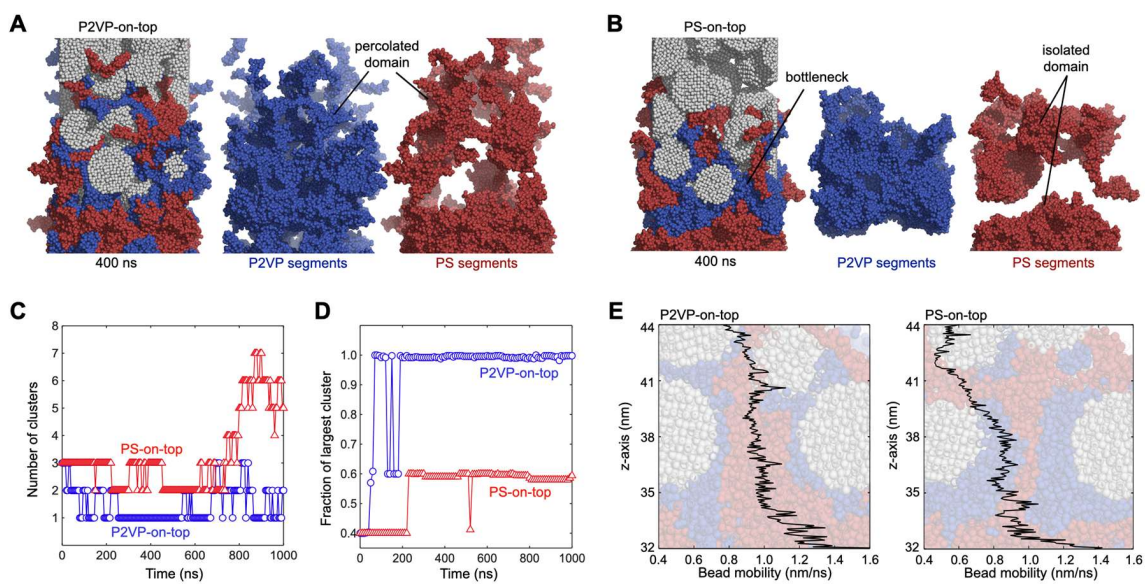

**Figure 4.** Domain connectivity and cluster analysis for two types of lamellar PS-*b*-P2VP films. (*A* and *B*) Snapshots of the domain morphology at 400 ns for the P2VP-on-top (*A*) and PS-on-top (*B*) configurations. From left to right, each panel shows the full morphology, P2VP segments (blue), and PS segments (red). (C) Time evolution of the number of PS clusters for P2VP-on-top (blue circles) and PS-on-top (red triangles). (D) Time evolution of the fraction of the largest PS cluster, defined as the number of beads in the largest cluster divided by the total number of PS beads (60,000 PS beads in total system), for P2VP-on-top (blue circles) and PS-on-top (red triangles). (*E*) Bead mobility along the z-axis coordinate for P2VP-on-top and PS-on-top systems.

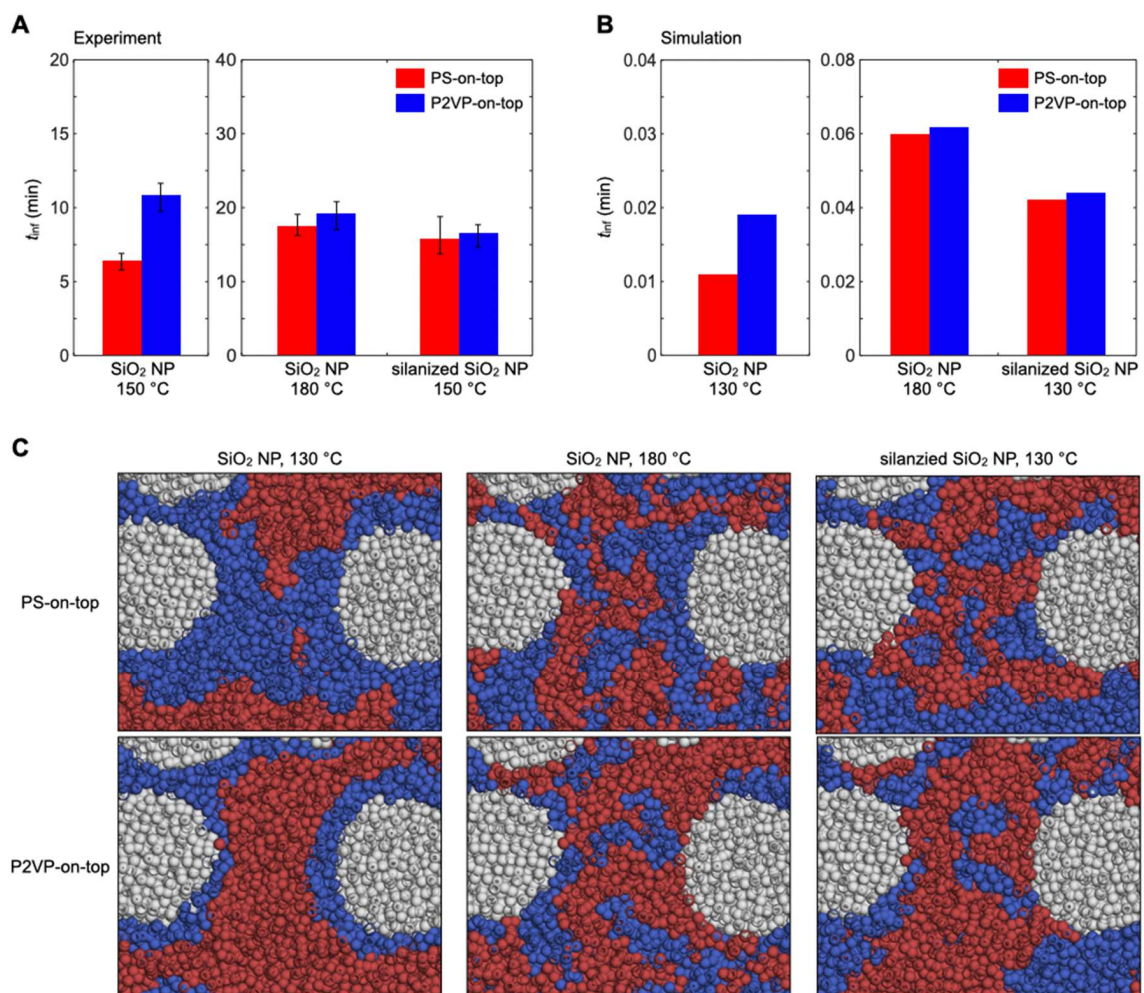

**Figure 5.** Influence of temperature and NP surface passivation on infiltration kinetics. (A) Experimental infiltration times for PS-on-top (red) and P2VP-on-top (blue) films at temperatures 150 °C (below $T_{ODT}$), 150 °C (above $T_{ODT}$) and for infiltration into silanized SiO$_2$ NP packings at 150 °C. (B) Simulated average infiltration rates, calculated from 200–800 ns, for PS-on-top (red) and P2VP-on-top (blue) at 130 °C (below $T_{ODT}$), 180 °C (above $T_{ODT}$) on SiO$_2$ NPs and 130 °C for silanized NP. (C) Representative spatial distributions of segments around nanoparticles: SiO$_2$ NPs at 130 °C (left), SiO$_2$ NPs at 180 °C (center), and silanized NPs at 130 °C (right). P2VP segments are shown in blue, PS segments in red, and nanoparticles are shown as grey spheres.

**Table 1.** Polymer Characteristics.

| Polymer | $M_{n,NMR}{}^{a}$ [kg/mol] | $M_{nSEC}{}^{b}$ [kg/mol] | Đ | $f_{PS}{}^{c}$ | $f_{P2VP}{}^{c}$ | $T_g{}^{d}$ [°C] | $T_{ODT}{}^{d}$ [°C] | CR[e] |
|---|---|---|---|---|---|---|---|---|
| PS-*b*-P2VP | 16.2 | 12.2 | 1.19 | 0.52 | 0.48 | 96 | 167 | 1.03 |

[a] Determined by ¹H NMR analyses in CDCl₃ (see SI Appendix)

[b] Determined by SEC analyses in THF (see SI Appendix)

[c] Mole fraction of PS and P2VP blocks determined by ¹H NMR analyses (see SI Appendix)

[d] Obtained from in-situ spectroscopic ellipsometry

[e] Confinement ratio (=Radius of gyration ($R_g$)/$R_{pore}$)